# The Trust Calibration Maturity Model for Characterizing and Communicating Trustworthiness of AI Systems

Authors: Scott Steinmetz[1] (ststein@sandia.gov, ORC), Asmeret Naugle[1] (ORC), Paul Schutte[1] (ORC), Matt Sweitzer[1] (ORC), Alex Washburne[1] (ORC), Lisa Linville[1] (ORC), Daniel Krofcheck[1] (ORC), Michal Kucer[2], Samuel Myren[2,3] (ORC)

1. Sandia National Labs, Albuquerque, NM, USA
2. Los Alamos National Labs, Los Alamos, NM, USA
3. Virginia Tech, Blacksburg, VA, USA

**Abstract:** Recent proliferation of powerful AI systems has created a strong need for capabilities that help users to calibrate trust in those systems. As AI systems grow in size and complexity, information required to evaluate their trustworthiness becomes less accessible, presenting a growing risk of using these systems inappropriately. We propose the Trust Calibration Maturity Model (TCMM) to characterize and communicate information about AI system trustworthiness. The TCMM incorporates five dimensions of analytic maturity: Performance Characterization, Bias & Robustness Quantification, Transparency, Safety & Security, and Usability. The TCMM can be presented along with system performance information to (1) help a user to appropriately calibrate trust, (2) establish requirements and track progress, and (3) identify research needs. Here, we discuss the TCMM and demonstrate it on two target tasks: using ChatGPT for high-consequence nuclear science determinations and using PhaseNet (an ensemble of seismic models) for categorizing sources of seismic events.

**Acknowledgement:**
Sandia National Laboratories is a multi-mission laboratory managed and operated by National Technology & Engineering Solutions of Sandia, LLC (NTESS), a wholly owned subsidiary of Honeywell International Inc., for the U.S. Department of Energy's National Nuclear Security Administration (DOE/NNSA) under contract DE-NA0003525. This written work is authored by an employee of NTESS. The employee, not NTESS, owns the right, title and interest in and to the written work and is responsible for its contents. Any subjective views or opinions that might be expressed in the written work do not necessarily represent the views of the U.S. Government. The publisher acknowledges that the U.S. Government retains a non-exclusive, paid-up, irrevocable, world-wide license to publish or reproduce the published form of this written work or allow others to do so, for U.S. Government purposes. The DOE will provide public access to results of federally sponsored research in accordance with the DOE Public Access Plan.

## Introduction

Rapid advances in AI technology, including foundation models and generative AI, have led to the proliferation of powerful new systems. However, the increasing size and complexity of these systems makes it challenging to gather information about their trustworthiness. Users need trustworthiness information to understand when and how AI systems should be used, while developers can use trustworthiness information to identify development needs. This article presents a Trust Calibration Maturity Model (TCMM) designed to characterize and communicate information about the trustworthiness of an AI system for use on a target task along five dimensions of analytic maturity: performance characterization, bias & robustness quantification, transparency, safety & security, and usability.

Calibrating trust in AI can be complex. Trust depends on risk; for example, a user with a low-risk task might be willing to trust a model relatively blindly, while users completing high-risk tasks would not. Users are vulnerable to the possibility that the AI system's results are incorrect or not aligned with user goals, so trust in the system depends on expectations of system performance (Mayer et al., 1995). Proper calibration of trust is crucial; a user should trust a model as much as it deserves, but not more (Jacovi et al., 2021). Trust is also dynamic, changing with the user's experience and the system's reputation (*Human-AI Teaming*, 2022; Miller, 2021).

The TCMM strives to bridge information on the *trustworthiness* of an AI system with information required for a user to calibrate their *trust* in the system for an intended *Target Task* (Figure 1). An *AI System* might perform multiple *System Tasks*. For example, a large language model (LLM) (the AI System) can be used for summarization, idea generation, question answering, etc. (the System Tasks). The user intends to complete some *User Task*. The TCMM characterizes the trustworthiness of the AI System in relation to the *Target Task*, which leverages System Tasks to support the User Task. A fully mature system characterization (an aspirational concept) would provide the user with all necessary information to calibrate their trust in the AI System for performing the Target Task. Different Target Tasks can require different levels of maturity across TCMM dimensions; for example, an analyst studying high risk national security applications needs to be able to trust the AI system they're using more than a curious student.

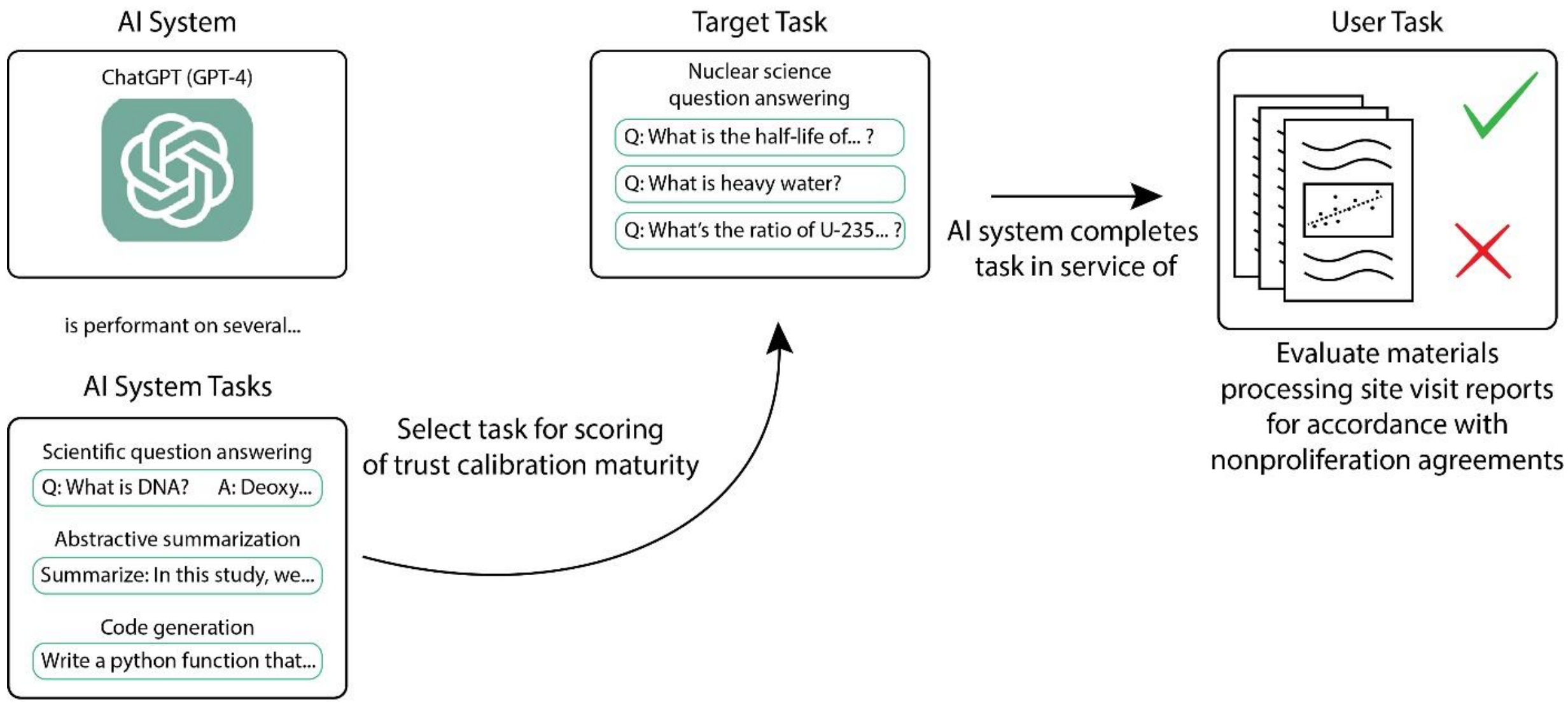

*Figure 1 - The TCMM characterizes the maturity of an AI System designed to accomplish a specific Target Task selected from a set of System Tasks that helps the end user accomplish the User Task. This hypothetical example focuses on nuclear science question answering as the Target Task in service of an end user evaluating a materials processing site visit report for accordance with nonproliferation agreements.*

This article introduces a Trust Calibration Framework (TCF) that describes factors contributing to trust calibration as well as the TCMM which scores the maturity of these factors for a given system. We also demonstrate use of the TCMM on two AI systems and target tasks.

## AI Trust Frameworks and other Maturity Models

While other frameworks and maturity models have been proposed for understanding and evaluating the trustworthiness of AI systems, existing options are not widely used and are generally not targeted toward use on high-consequence applications. Some of these reasons are simply the inertia of the field or limited knowledge of these frameworks, but others are intrinsic to the design of the maturity models. Extant trust in AI system frameworks are discussed below in The Trust Calibration Framework section; they describe aspects of AI systems that are important to Trust without providing a scoring system for evaluation of maturity. These existing frameworks are useful for guiding development of AI systems generally, but none met our needs of (1) emphasizing factors that are relevant to modern, foundation model-based AI systems and high-consequence applications and (2) being succinct enough to serve as useful communication tools.

Schaschek & Engel (2023) introduce a maturity model focused on the AI design process, specifically focusing on data, technology, people & culture, and processes. This model targets management, societal, and organizational perspectives on trust, including ethics and fairness and looks at the maturity of an approach. This is significantly different from the TCMM, which targets expert users and focuses on the maturity of an AI system and its suitability for a target task. Cho et al. (2023) and Das et al. (2024) also introduce AI maturity models. Cho et al. focus on the AI development process, but do not include specific maturity requirements for trustworthiness dimensions evaluated on a particular tool as we do with the TCMM. Das et al.'s maturity model is the most similar of existing models to the TCMM. The model focuses on fairness, explainability, auditability, and safety, and identifies four possible overarching maturity dimensions. We focus on maturity dimensions that are targeted more toward utility for expert users, although are dimensions are also applicable to non-experts. Das et al. also stop at high-level maturity descriptions, while we define specific requirements for reaching each maturity level in each dimension.

We also find that previous maturity models failed to break out a core difficulty of modern AI: system characteristics must be well-understood in relation to the goal the AI system is intended to achieve (i.e. the target and user task). The ambiguity of modern measures of performance, bias & robustness, and transparency & explainability on complex, subjective tasks typically performed by humans underlies much of the mistrust users of AI systems currently have. How does one calibrate their trust that an AI System will appropriately answer nuclear science questions when those could be open-ended? When an uncountable number of variations on the correct (or incorrect) answer could be given? When even an answer produced by an expert human requires significant vetting or trust? Unlike many problems solved by traditional methods, available performance metrics are often only coarse proxies for the true task performance. For instance, comparison of model answers to general science benchmarks using lexical similarity metrics like BLEU and ROUGE can capture how well a model has produced those *exact answers* but is insufficient for equating the semantic meaning of differently worded correct answers. Further, such a benchmark cannot capture performance on any field *not* within that benchmark. This is a fundamental limitation of all measures, but AI systems often suffer drastic, performance losses in out of distribution domains, even those that seem related to human users. Many models which score well on benchmarks suffer from hallucinations and context issues in practice, preventing adoption for specific applications. Despite these limitations, general benchmark performance is often the only available measure when touting AI system capability.

## The Trust Calibration Framework

We identify the appropriate dimensions of analytic maturity to include in the TCMM with the introduction of the Trust Calibration Framework (TCF), which outlines factors that determine user trust in an AI system (Figure 2). While the TCMM is intended for broad use on diverse AI systems and target tasks, we emphasize factors that are relevant to modern, foundation model-based AI systems and high-consequence applications; we did not find existing frameworks that were both relevant to this combination and succinct enough to serve as useful communication tools. A trust framework that cannot be easily communicated to nonexperts (e.g. within a single hour-long meeting) faces significant barriers to adoption and use. The five factors of the TCF incorporated as dimensions in the TCMM are highlighted in yellow in Figure 2 and described in more detail later in this article.

The TCF begins by considering user involvement in the development of the AI system, which includes system design, data selection/collection, model training, and development of tools that comprise the AI system. Involving target end users in the system design process can improve the system's alignment with user needs (Kujala, 2003), leading to a better user experience (Baroudi et al., 1986) and improved user trust (Bano & Zowghi, 2015). Characteristics of the AI system itself that contribute to system trustworthiness include performance characterization, bias & robustness quantification, transparency, and safety & security. These factors combine with usability to determine the quality of the user's experience with the AI system, which contributes to user trust. This experience is dynamic; as users interact with an AI system their opinion about it may change. The system's reputation, which can be shaped by reviews from other users, also affects user trust (AI HLEG, 2019; Corritore et al., 2003; Dorton & Harper, 2022).

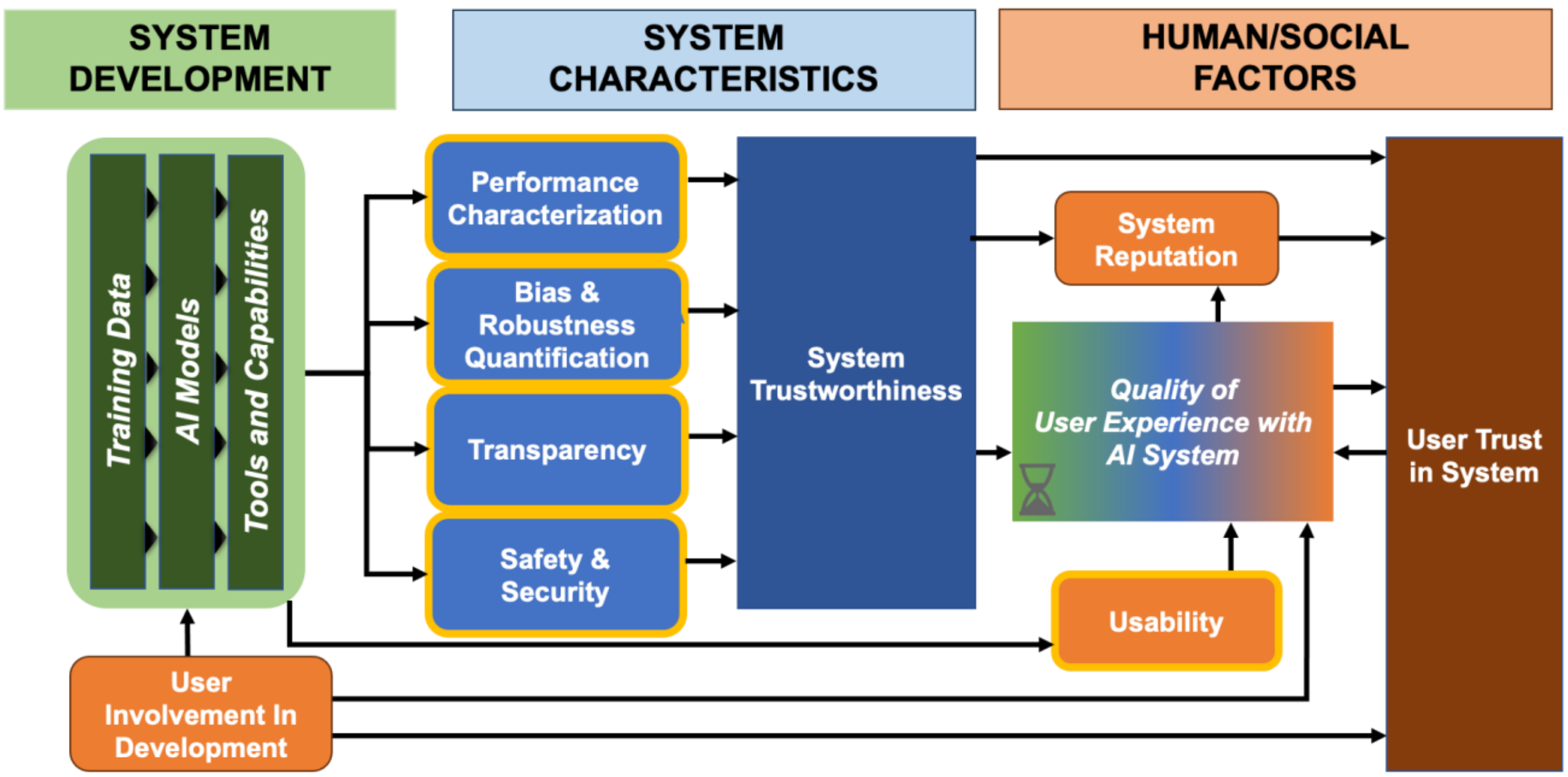

*Figure 1: The Trust Calibration Framework (TCF) describes relationships between factors that contribute to user trust in an AI system; elements included as dimensions in the Trust Calibration Maturity Model (TCMM) are highlighted in gold.*

To develop the TCF, we surveyed existing trust models in the literature (AI HLEG, 2019; Buchheit et al., 2021; Chiou & Lee, 2023; Corritore et al., 2003; Dorton & Harper, 2022; Hancock et al., 2011; Hoff & Bashir, 2015; Mayer et al., 1995; Sheridan, 2019; Speed & Stracuzzi, 2020; Stanton & Jensen, 2021; *Trustworthy Artificial Intelligence (AI)™*, 2020) and consolidated like factors (Table 1). We found that many factors mentioned in the literature overlap or represent similar concepts. In these cases, we strove to identify a dimension that captured the character of the set of similar factors. For example, characterizing the performance of an AI system can include metrics of its reliability, accuracy, consistency, competency, integrity, dependability, maturity, quality, responsibility, and variability. For some applications it may be important to separate those concepts. For the TCMM, our goal is to characterize the maturity of analytic efforts generally, so we chose to consolidate these concepts into a performance characterization dimension. For two of our dimensions, we combined distinct factors with significant conceptual overlap. We found that bias and robustness play similar roles in determining user trust, with bias in system inputs heavily related to system robustness. We also chose to consolidate safety and security, as the concepts are both related to protecting against threats. Even with this consolidation, TCMM dimensions are not independent. For example, performance impacts robustness. Nevertheless, we selected these dimensions as the most descriptive set that we felt was parsimonious enough to effectively communicate trustworthiness information to experts and non-experts alike.

*Table 1 Relationship between TCMM dimensions and factors discussed in surveyed trust models*

| TCF/TCMM Dimension) | Factors in Literature Incorporated (number of surveyed frameworks that include factor) |
|---|---|
| Performance Characterization | Reliability (7), Accuracy (4), Consistency (4), Performance (4), Competency (3), Integrity (3), Dependability (2), Maturity (2), Quality (2), Responsibility (2), Variability (1) |
| Bias & Robustness Quantification | Reliability (7), Context (5), Performance (4), Resilience (3), Robustness (3), Dependability (2), Fairness (2), Vulnerability (2), Variability (1) |
| Transparency | Transparency (7), Consistency (4), Explainability (4), Predictability (4), Auditability (3), Variability (1) |
| Safety & Security | Accountability (4), Privacy (4), Risk (4), Security (4), Safety (3), Compliance (2), Dependability (2), Responsibility (2), Vulnerability (2) |
| Usability | Transparency (7), Consistency (4), Usability (4), Acceptability (1) |
| | *Not aligned with TCMM dimensions: Reputation (3), Situation (3), Suitability (2), Sustainability (2), Assurance (1), Maintainability (1)* |

## The Trust Calibration Maturity Model

The TCMM (Table 2) characterizes and communicates information about the maturity of system trustworthiness. It is applied to an AI system used on a specified target task, and considers five dimensions (performance characterization, bias & robustness quantification, transparency, safety & security, and usability) identified in the TCF. For each dimension, the TCMM defines criteria for four levels of analytic maturity. The TCMM was inspired by previous maturity models such as Capability Maturity Models (Paulk et al., 1993), Technology Readiness Levels (Mankins, 1995), and the Predictive Capability Maturity Model (Oberkampf et al., 2007). In the TCMM, a dimension that has not been addressed is assessed to have level 1 maturity. Level 2 maturity means that basic steps have been taken to characterize system trustworthiness along that dimension. Level 3 involves systematic analysis, and level 4 incorporates both comprehensive trustworthiness characterization and consideration of the human-AI team. Most dimension-level pairs specify multiple criteria; in these cases, all listed criteria must be met to reach that level of maturity. When producing a TCMM score for a given AI system and User Task, scorers specify which system components or metrics satisfy the criteria for that maturity.

Maturity requirements vary across AI systems and target tasks, often not requiring maximum maturity. Level 4 maturity might be considered aspirational for many current AI systems. Our intent is to make system maturity explicit, facilitating structured discussion around trust calibration and development decisions, and the TCMM can be used flexibly depending on the use case. Three primary use cases are: (1) communicating system trustworthiness to users, (2) specifying and tracking progress toward trustworthiness requirements during development, and (3) specifying techniques available for reaching different maturity levels to highlight gaps and research needs.

| Dimension of Trustworthiness Assessment | Level 1 (not addressed) | Level 2 (basic) | Level 3 (systematic) | Level 4 (comprehensive) |
|---|---|---|---|---|
| **Performance Characterization** | Performance is unknown or unknowable. | Performance on at least one system task similar to the target task is known. | Performance on the target task is systematically evaluated.<br><br>At least one metric of uncertainty relevant to the target task has been estimated. | Performance for all dimensions relevant to the target task is known.<br><br>Metrics of uncertainty relevant to the target task, and sources of that uncertainty, have been estimated with known confidence.<br><br>End user performance on the user task is measured. |
| **Bias & Robustness Quantification** | Bias and robustness are not measured or considered. | Major sources of potential bias have been identified.<br><br>Bias is measured along at least one relevant dimension.<br><br>At least one form of robustness is measured. | Multiple known relevant dimensions of bias and robustness are systematically measured for the target task.<br><br>System communicates known sources and magnitudes of bias to user.<br><br>Domain and conditions in which system should (and should not) be used are communicated to the user. | All known relevant dimensions of bias and robustness are comprehensively and systematically measured.<br><br>Results estimate error contributed by known sources of bias.<br><br>System provides a mechanism for human-AI team to proactively identify new conditions and sources of bias, and to test for robustness to those.<br><br>Retraining criteria are defined. |
| **Transparency** | Mapping between system inputs and outputs is unknown or unknowable; unexpected results are common.<br><br>Mechanism of system function (e.g. model weights) is inaccessible. | Users can infer a coarse mental model of how the system produced an output; unexpected results occur in core capabilities.<br><br>Mechanism of system function is interrogated anecdotally. | Users maintain a faithful but approximate mental model of the system; unexpected results occur in some cases, usually in less-core capabilities.<br><br>Mechanism of system function is interrogated systematically and communicated to the user. | Users have an accurate mental model of the system and how results were reached; users are rarely unable to understand system results after investigation.<br><br>Rationale underlying system outputs is known and communicated to user. |
| **Safety & Security** | No user guardrails are in place.<br><br>System security is not considered. | User guardrails exist.<br><br>Counter-adversarial measures are implemented by third parties but unverifiable by user.<br><br>User inputs and system responses are accessible to third parties. | System prevents both direct and indirect efforts to misuse, and has been tested through red teaming.<br><br>Some counter-adversarial measures are implemented by trusted parties.<br><br>User inputs and system responses are accessible only to trusted parties.<br><br>Training data sources are known and reported to user. | System prevents sophisticated misuse efforts, as confirmed through red teaming.<br><br>Counter-adversarial measures account for all known threats, evolve with new threats, and are implemented by trusted parties.<br><br>User inputs and system responses are accessible only to task-specific trusted parties.<br><br>System is trained on known and controlled datasets only. |
| **Usability** | Interface(s) either does not exist or is targeted toward developers. | User interface(s) exists.<br><br>Basic usability features are incorporated, and some user/task needs are met.<br><br>User training or documentation exist. | User interface(s) has undergone established usability testing and incorporated the results as appropriate.<br><br>User training and documentation are comprehensive.<br><br>User feedback is collected and considered. | Interface(s) is intuitive, adaptive, and tailored to individual analyst preferences.<br><br>User training and documentation are well-targeted to task.<br><br>User feedback is collected and utilized for continuous improvement. |

*Table 2: The Trust Calibration Maturity Model (TCMM) provides a structured path for characterizing and communicating the maturity of trustworthiness of a given AI system for use on a target task.*

## Dimensions of Trust Calibration Maturity

### Performance Characterization

Performance, the ability of an AI system to do its task well, is a major contributor to user trust (Dorton & Harper, 2022). Performance can involve the system's competence, accuracy, errors, and reliability, along with quantification of confidence and uncertainty. Since even high-performing systems are not always correct, the TCMM captures how well an AI system's performance is *known*. High maturity in performance characterization indicates that we are confident in our understanding of the system's performance, not necessarily that the performance itself is high. Performance requirements are determined by the user and target task; for example, in high-consequence settings the user might tolerate more false alarms to avoid missing a single true positive.

Relevant metrics of performance heavily depend on the systems and target tasks being evaluated. These might include traditional machine learning evaluation metrics, such as classification accuracy, logarithmic loss, confusion matrices, area under the curve, F1 score, and mean absolute or squared error (Mishra, 2018). Metrics for evaluating large systems vary, for LLMs these can include exact-match accuracy, F1 score for word overlap, MRR and NDCG scores, ROUGE score, and many more (see (Liang et al., 2023) for a discussion of holistic metric usage). For foundation models and generative AI, performance is often measured against public benchmarks on broad tasks related to the application the system is designed for. The usefulness of such benchmarks is based on the underlying assumption that benchmark performance will be informative about performance on various specific applications, though this varies in practice. Performance metrics might also measure reliability, reproducibility, and uncertainty, with specific metrics dependent on the target task, or might focus on the performance of the human-AI team on the user task (Figure 1).

Level 1 performance maturity means that performance is unknown. Level 2 is reached if at least one performance metric has been characterized for a *system* task with some expected relevance to the target task. Level 3 requires that performance be measured systematically on the target task, and with some uncertainty quantification (UQ). Level 4 requires mature performance characterization along all known dimensions relevant to the target task, comprehensive UQ with known confidence, and known performance on the user task tackled by the human-AI team.

### Bias & Robustness Quantification

Bias in an AI system involves systematic error that results in inconsistent model performance. While all models are biased, we focus on assessing whether an AI system's bias is well characterized and communicated to the user. If unknown to the user, bias can result in poorly calibrated trust and misapplication of the AI system. Algorithmic bias is bias inherent in the design and training of a model, and can stem from sources such as feature selection, model architecture, and hyperparameter choices. Data bias refers to bias in the training data, and can be a result of data collection methods, sample selection, or historical biases (Tustison et al., 2013) that are ultimately encoded in the AI system. Output bias occurs when model products incorporate data or algorithmic bias. The negative impacts of bias may be compounded with human-AI interaction, as users are prone to their own automation and complacency biases when the system's reasoning processes are obscure (Goddard et al., 2012) or when statistical confidence is high (Parasuraman & Manzey, 2010).

Robustness is the stability of key performance indicators (KPIs) to changes in the training data, input data, model, or application space. Robustness metrics can quantify the reliability of a model's desired function under novel applications, adversarial manipulation, anticipated changes to data streams, and other real-world settings not necessarily represented during training. Robustness quantification requires specification of KPIs, tolerance for changes in the KPI, and specific changes in the data, model, or application space relative to those used during model training.

Bias and robustness are fundamentally intertwined, and thus paired in the TCMM. Bias can cause inconsistent performance, leading to low robustness across applications. Maturity levels begin with level 1, in which bias and robustness are not considered. Level 2 requires that major sources of bias are identified, and that bias and robustness are each measured along at least one dimension relevant to the system task. Level 3 requires that bias and robustness be systematically measured along multiple relevant dimensions, known sources and magnitudes of bias are characterized and communicated to the user, and the user be made aware of the conditions under which the model performs well. Level 4 involves systematically measuring all known sources of bias and robustness, estimating the error that these cause, and a mechanism for the human-AI team to analyze new bias and robustness concerns with criteria for model retraining.

### Transparency

Transparency involves the ability of a person to understand and predict the AI system's performance, intent, and reasoning processes (*Human-AI Teaming*, 2022). Explainability and interpretability fall into this dimension, despite some definitions noting technical differences (Tabassi, 2023). Transparency can help users to develop more accurate mental models of the AI system, better understanding how it produces its output. Since transparency depends on user comprehension, the targeted userbase is key in determining requirements and measuring effectiveness. There are at least two audiences for transparency efforts: AI developers can use their understanding of model function to improve the system, while end users require transparency to calibrate trust.

Available transparency methods depend on the AI system and modalities involved. Many modern AI systems are not interpretable, and LLMs in particular often rely on post-hoc explanations with limited faithfulness or fidelity to model processes (Bubeck et al., 2023; Liao & Vaughan, 2023; Papenmeier et al., 2019; M. R. Smith et al., 2022). The lack of transparency in many AI systems limits their applicability to many real-world domains, with immature and incorrect explanations negatively impacting decision-making (Stites et al., 2021). LLMs can utilize methods like chain of thought prompting to clarify reasoning strategies (Wei et al., 2022) which does not provide true mechanistic insight but enables human users to verify the ostensible underlying logic of a response. Few current techniques provide true mechanistic insight, but this is active research with several promising methods such as sparse autoencoding dictionaries identifying monosemantic neural circuits (Templeton, 2024).

At TCMM Level 1, transparency is not considered. Level 2 requires that users be able to coarsely understand how the system produces results, and that developers be able to interrogate system process anecdotally. At Level 3, users can develop faithful but approximate mental models (Hoffman et al., 2019) of the system, and developers can systematically interrogate system response generation. Level 4 indicates that users can develop *accurate* mental models of the system and developers have a clear understanding of the response generation process. Because system transparency is framed around the impact on user understanding of the system, a fully mature (Level 4) measurement of transparency requires end user testing.

### Safety & Security

Trust in an AI system requires consideration of the risks associated with its use. AI system safety, which focuses on preventing harm *from* the system, and security, which focuses on preventing harm *to* the system, contribute to this risk (Qi et al., 2024). Safety includes efforts to prevent misuse of AI systems, including malicious use (Hendrycks et al., 2023) and negative side effects (Amodei et al., 2016) that may or may not overlap with misuse of automation tools in general (Parasuraman & Riley, 1997). Security aims to prevent adversaries from accessing or tampering with the AI system. Security includes consideration of many potential types of attacks, including attacks related to data collection (e.g. data bias, fake data, data breach), poisoning attacks (gradient-based, clean label poisoning, backdoor attack, GAN-based attacks, e.g. (Gu et al., 2019)), adversarial attacks (white-box, black-box, gray-box attacks, image-scaling attacks, e.g. (Goodfellow et al., 2015; Quiring et al., 2020)), and system integration attacks (AI confidentiality, AI bias, code vulnerability, e.g. (Hu et al., 2021; Radosevich & Halloran, 2025; M. Smith et al., 2023)), as well as inference and reconstruction attacks, model stealing attacks, and resource-depletion attacks (Bommasani et al., 2022). Further, inaccessibility of training data and training approaches can hamper inspection or awareness of vulnerabilities. Recent shifts toward foundation models have increased the threat landscape, primarily because most foundation models are developed and controlled by third party commercial entities as a service, sending user interactions to provider computing systems and creating confidentiality concerns.

The target task determines requirements regarding AI safety and security. Publicly available systems may require significant safety consideration to ensure that they are not misused (e.g. dual-use risk), while high-consequence national security applications may place more emphasis on security: assurance that a system is secure from external tampering. System security can be addressed through mechanisms such as authentication, validation, red teaming, and firewalling and isolation. System safety is often addressed through 'guardrails' that limit use to vetted and appropriate lanes. Guardrails may range from alignment to safe answers during model training, to general warnings about the safe use of the system, to automated warnings identifying unsafe or insecure activities, to enforced prohibitions on system responses of concern.

Maturity levels begin with level 1, in which safety and security are not considered. Level 2 requires that some user guardrails are implemented to prevent misuse, and that some security measures have been taken although their details may not be known to the user. Further, the system may not keep user information or activity confidential. Level 3 requires strong guardrails, security measures implemented by trusted parties, confidentiality of user inputs and system responses, and known sources of training data. At level 4, even sophisticated attempts at misuse are prevented and tested through red teaming. All known security threats are managed by trusted parties, and new threats are identified and neutralized., confidentiality is closely maintained, and training datasets are controlled.

### Usability

Usability refers to the quality of a user's experience when interacting a system, or “the extent to which a system, product, or service can be used by specified users to achieve specified goals with effectiveness, efficiency, and satisfaction in a specified context of use” (International Organization for Standardization (ISO), 2018). Elements of usability include intuitiveness, ease of learning, efficiency, memorability, and error proneness (Nielsen, 1994). Poor usability can lead to human error and a reduction in efficiency (Goodwin, 1987), whereas good usability can reduce training, reduce errors, and improve performance.

Usability cannot readily be assessed until a user interface and workflow have been created, so usability concerns generally become important at Technology Readiness Levels (Mankins, 1995) 4 and higher. There are several methods available to assess the usability of a system and its interface. Method selection depends on the type of system and target task. Usability testing involves human assessment but does not require large numbers of subjects (Nielsen, 2012), generally involving a battery of tests that assess key usability metrics.

At Level 1, no user interface exists; any interface is targeted only toward developers. Level 2 requires a basic user interface with some usability features that meet some user task needs, as well as training or documentation. At Level 3, the system and its interface are fully tested for usability and refined based on that testing. Documentation and training are comprehensive and user feedback is considered. A level 4 system tailors its user interface to the individual user, targets training and documentation toward the user task, and includes mechanisms for continuous improvement.

## How to Use the TCMM: Scoring Process and Examples

The TCMM should be scored by an expert familiar with the AI system. Once the user and target tasks are identified, each dimension of the TCMM is scored independently. Reaching a maturity level for a particular dimension requires the system to meet *all* criteria in that level; only the lowest level entirely satisfied should be assigned as the final score, though individual criteria that meet higher maturity levels can be noted. When producing a TCMM score for a given AI system and User Task, scorers should specify which system components or metrics satisfy the criteria for the scored level.

To illustrate the scoring process, we give two examples.

### Example 1: ChatGPT for Nuclear Science Question Answering, no web-access

Consider a hypothetical analyst evaluating site visit reports to assess adherence to international nuclear nonproliferation agreements (Figure 3). This is a high-consequence task that demands attribution and clear explanations for any conclusions drawn. The analyst has access to an instance of ChatGPT without live access to the internet to help with nuclear science questions (e.g. Table 3) that arise while assessing the report. They use the system to accelerate their assessment but must not over-rely on any individual answer. Many analysts and teams are (appropriately) reluctant to use AI systems for such use cases; AI used for this application must bridge the trust gap.

**AI System:** ChatGPT (GPT-4-Turbo) without access to the web
**User task:** nonproliferation analyst evaluating reports to assess accordance with nonproliferation agreements
**Target task:** nuclear science question answering

| Question | Answer | Difficulty | Type |
|---|---|---|---|
| What is the cutoff energy for a thermal neutron? | 0.025 eV | Easy | Short QA |
| How many protons, neutrons, and electrons does a calcium-40 isotope have if its charge state is 2+? | 20 protons, 20 neutrons, 18 electrons | Medium | Short QA |
| Why can't Geiger-Muller counters be used for spectral analysis? | The electric field near the anode wire is so strong that the gas multiplication factor is extremely high. This means that when an event happens, so many electrons are created during a successful avalanche that the electronics can't tell the difference between particles. | Hard | Open QA |

*Table 3: Example nuclear science questions from the 100 question NuclearQA dataset (Acharya et al., 2023). Questions can be easy, medium, or hard difficulty and labeled answers are either short or open-ended. For the purposes of this scoring example, these questions stand as good proxies for those an analyst might ask when assessing site visit reports for congruence with nuclear nonproliferation agreements.*

| | Level 1 | Level 2 | Level 3 | Level 4 |
|---|---|---|---|---|
| **Performance Characterization** | | ScienceQA – 32nd<br>Chatbot arena – 50th | | |
| **Bias & Robustness Quantification** | | Warnings to user<br>Chatbot Arena – 7th | | |
| **Transparency** | Private weights | Post-hoc explanations | | |
| **Safety & Security** | | Inputs sent to 3rd party<br>Private training data | Advanced guardrails<br>Tested via red teaming | |
| **Usability** | | | Intuitive interface<br>Customizable GPTs | |

*Figure 2: TCMM rating for ChatGPT 4 for nuclear science question answering. Logo indicates TCMM score. Text below the logo denotes benchmark rankings or system components that support the score. Text in a higher level than the score indicates individual criteria that satisfied higher maturity levels.*

**Performance characterization: Level 2.** ChatGPT's performance has been evaluated on related public benchmarks such as ScienceQA (Lu et al., 2022) and Chatbot Arena (Chiang et al., 2024), but not benchmarks specific to the target task of *nuclear* scientific question answering. Testing on NuclearQA, a 100 question dataset (Acharya et al., 2023) directly relevant to the target task, provides some measure of target task performance, though is small in scale. ChatGPT does not provide uncertainty metrics to a user asking scientific questions.

**Bias and robustness quantification: Level 2.** As with performance, bias and robustness have been explored (Chiang et al., 2024; Ray, 2023; Wang et al., 2023) but not for applications directly relevant to the target task. ChatGPT provides some textual warnings to the user about specific biases (data bias, final date of training, etc).

**Transparency: Level 1.** ChatGPT can provide post-hoc explanations when prompted but users cannot access the model weights, so interpretability techniques requiring model weights are not possible. This internet-restricted version of ChatGPT cannot provide citations for specific claims.

**Safety & Security: Level 2.** ChatGPT has guardrails to thwart attempted system misuse (*How to Implement LLM Guardrails | OpenAI Cookbook*, n.d.; *How to Implement LLM Guardrails | OpenAI Cookbook*, n.d.; *OpenAI Safety Practices*, 2024). OpenAI holds red teaming exercises and

issues reports on safety and misuse risks (*Lessons Learned on Language Model Safety and Misuse*, 2024). However, full details on OpenAI's vetting processes are not available to the user. Use of ChatGPT involves sending user inputs externally to OpenAI, a non-trusted third party in this example.

**Usability: Level 3.** ChatGPT has an intuitive interface well-suited to the target task, with useful documentation, opportunity for user customization, and user feedback.

### Notes Regarding Trust for Target User and Task

ChatGPT is widely regarded as one of the leading AI systems with broad public appeal. A lack of benchmark options directly relevant to the target task keeps it from scoring higher on performance characterization. Despite relatively high usability, transparency and explainability may be a major barrier to calibrated trust for this target task.

## Example 2: Ensemble of PhaseNet models for Phase Picking

Consider an analyst working to characterize and localize seismic events to determine if they are due to earthquakes or mining explosions (Figure 4). Phase picking is a fundamental step in this process, where the analyst identifies when the primary (P) and secondary (S) phases of a seismic wave were sensed. The analyst is provided with an AI system (Myren et al., 2025) using an ensemble of models based on PhaseNet (Zhu & Beroza, 2018), a U-Net based (Ronneberger et al., 2015) deep learning model for seismic phase picking, trained on INSTANCE (Michelini et al., 2021) data from Italy, and with uncertainty incorporated through ensembling. The AI system intakes a tri-axial seismic waveform of size 3x3,000 (e.g. 30 seconds at 100 Hz in the vertical, north/south, and east/west directions) and outputs a matrix of pseudo-probabilities of the same size that are interpreted by the analyst to identify if and when the P and S waves arrive. An example of the AI System's input and output is provided in Figure XX.

**AI System:** Ensemble of PhaseNet models from (Myren et al., 2025)
**User task:** Seismic analyst performing seismic event detection and categorization
**Target task:** Offline phase picking for events in Italy

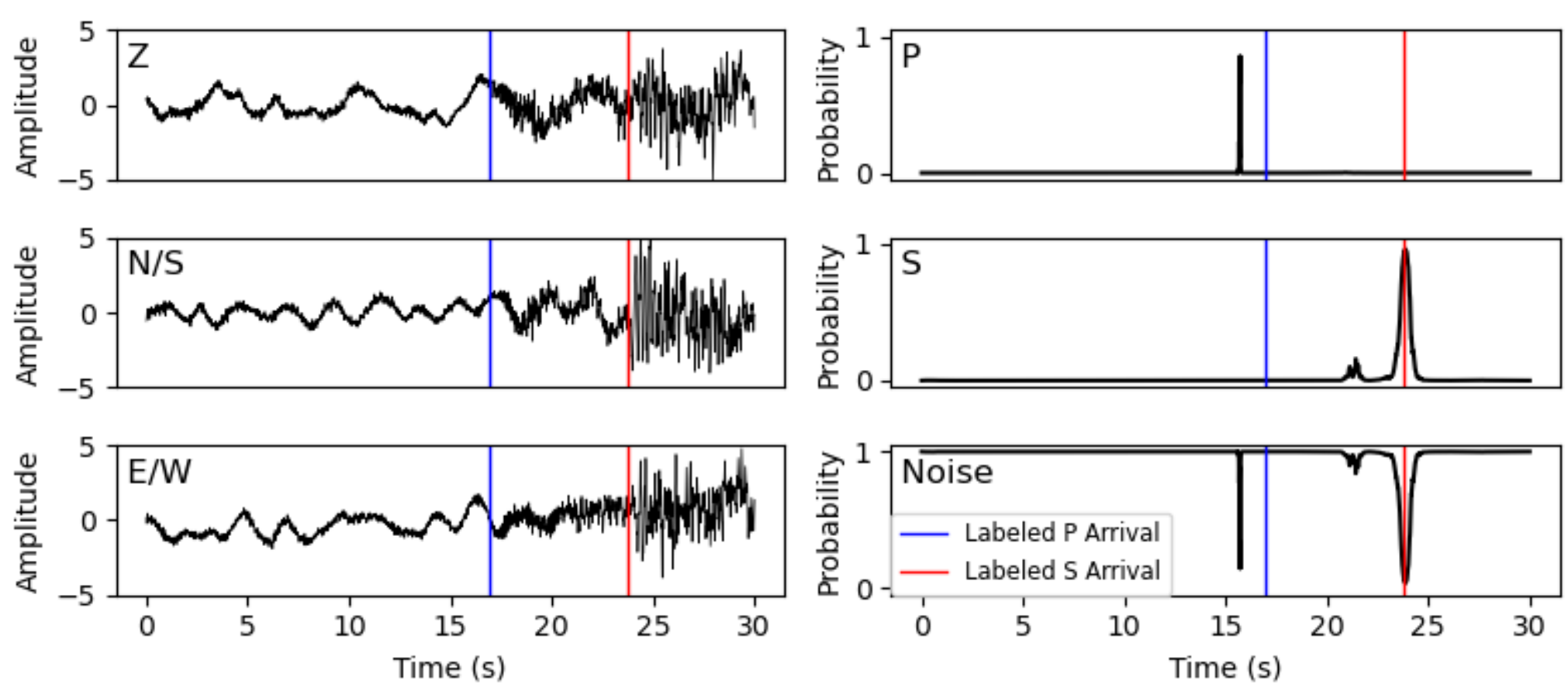

*Figure 3 - Shown is an example of the PhaseNet ensemble's input (left column) and output (right column). The input is a normalized, tri-axial seismic waveform of size 3x3000 (30 seconds sampled at 100 Hz in the vertical (Z; top left), north/south (N/S; middle left), and east/west (E/W; bottom left) directions. The output is a matrix of pseudo probabilities of the same size which suggest the probabilities of a P-wave arrival (P; top right), S-wave arrival (S; middle right), and noise (Noise; bottom right). The labeled P-wave arrival (blue) and S-wave arrival (red) are shown for reference, but these labels are unknown during test time. The analyst processes the AI system's outputs to predict if and when the P and S waves arrive when the labels are unknown.*

| | Level 1 | Level 2 | Level 3 | Level 4 |
|---|---|---|---|---|
| **Performance Characterization** | | | INSTANCE subset<br>Uncertainty via Ensembling | |
| **Bias & Robustness Quantification** | No robustness metrics | Bias sources identified<br>Data bias measured | | |
| **Transparency** | | Effortful causal probing | | |
| **Safety & Security** | No user guardrails<br>Security not considered | | Public training data (INSTANCE) | |
| **Usability** | Developer-facing model | SeisBench interface available | | |

*Figure 4 - TCMM rating for PhaseNet ensemble for phase picking. Logo indicates TCMM score. Text below the logo denotes benchmark rankings or system components that support the score. Text in a higher level than the score indicates individual criteria that satisfied higher maturity levels.*

**Performance Characterization: Level 3.** Performance metrics with ensemble-based uncertainty estimates are known for the target task itself on a related data set (Myren et al., 2025). For example, the Italian data can be expected to attain a recall metric of $0.80 \pm 0.05$ when trained as reported. Since the target task is seismic event detection in Italy, this dataset provides clear evidence of the AI system's performance on the target task.

**Bias and Robustness Quantification: Level 1.** While some sources of bias in the system are known (Chai et al., 2020; Münchmeyer et al., 2022; Myren et al., 2025; Park et al., 2024), no robustness estimates were calculated. Were a higher level desired, measures of the impact of noise or various forms of data augmentation would be a first step towards robustness metrics.

**Transparency: Level 2.** The system's transparency score of 2 reflects users' access to the model weights and training process which breaks out performance by region, allowing its target audience of expert users to develop coarse mental models through experiment.

**Safety and Security: Level 1.** No guardrails are in place, and security has not been considered.

**Usability: Level 1.** This is a analysis-oriented tool developed for research purposes and interacting with the model requires technical interaction with Python and PyTorch (Paszke et al., 2019). A light technical lift could make use of the SeisBench interface, lifting the AI system to a level 2 if desired.

### Notes Regarding Trust for Target User and Task

This PhaseNet ensemble system has well-characterized performance, but relatively low maturity on safety & security and usability. PhaseNet models are widely used for phase picking and event detection by seismologists due to their strong performance (Han et al., 2023; Zhu & Beroza, 2018).

## Limitations of the TCMM

One intrinsic limitation of the TCMM is that it is meant to be usable across many tasks and modalities, but must be task-specific to be meaningful. Because of this, future work may involve the development of more specific sub-criteria relevant to particular tasks or domains. This also means that a single AI system may need to be scored multiple (or even many) times to account for varied use cases.

Applying the TCMM is also somewhat subjective, and different evaluators may rate maturity differently. To evaluate the extent of this subjectivity, we conducted a small inter-rater reliability experiment for the ChatGPT Nuclear Science QA example to measure the impact of the subjective language and criteria on score consistency. Two independent raters with familiarity of the system were tasked with scoring on the five dimensions of the TCMM using Table 2 and no additional instructions. They produced the following ratings: Performance: 2, 2; Bias and Robustness: 2, 2; Transparency: 2, 1; Safety and Security: 3, 3; Usability: 3, 3. Combining these ratings with those presented above and using Cohen's $\kappa_w$ with a quadratic weight to penalize greater differences in scores, inter-rater reliability is 0.75, which is considered "excellent" agreement. While greater conceptual clarity of the Transparency dimension in the TCMM may have eliminated the single point of disagreement, we feel that additional explication of characteristics that qualify a model for a specific score may harm its generalizability to other applications. Importantly, this is also a single example demonstrating inter-rater reliability of the TCMM scoring system; a more robust assessment may aid in further clarifying the conceptual dimensions and values, but this is beyond the scope of this paper.

In practice, we find subjectivity in the level criteria is necessary to fit to real-world maturity scoring. Such subjective language (and scoring) also induces necessary conversation between developer and customer (or researcher and reviewer) around what the specific needs are for a target task and user task. We acknowledge that no one maturity model can fit all the various applications and contexts AI systems operate in yet offer the domain-agnostic TCMM as a useful starting point in such conversations. Some dimensions or specific level criteria may not require high levels of maturity for particular user and target task needs; the PhaseNet scoring above could be seen as one such example, where Safety & Security and Usability did not require high levels of maturity within the original research context.

## Conclusion

We present the TCF, which describes factors contributing to trust calibration, and the TCMM, which aims to provide a clear and actionable structure for characterizing and communicating information about the trustworthiness of AI systems. The TCMM can facilitate conversations about trust and trustworthiness that should take place between AI system developers and their users. We suggest that the TCMM be used to structure conversation around specific needs for the target task and the cost of meeting those needs. It provides guidance to developers who aim to build system trustworthiness and gives users the information they need to calibrate their trust appropriately.

The TCMM addresses shortcomings of existing models of trustworthiness .The TCMM is better suited than other Trustworthiness maturity models for addressing the complexity, difficulty, and subjectivity of many modern AI tasks AI by explicitly capturing the relationship between each dimension of maturity and the specific target and user tasks. The TCMM captures the maturity along dimensions held to drive user trust by the literature with the structure followed by modern AI R&D teams: development starts with general, related tasks (i.e. level 2 maturity), moves on to specific, often custom, measures of direct target task performance (i.e. level 3 maturity), and aspires to continual, online, and adaptive measures in deployment (i.e. level 4 maturity).

Application of the TCMM is specific to the target task, which means that an AI system may need to be scored for multiple uses. Additionally, we have designed the TCMM for broad applicability, leading to some subjectiveness in criteria. While no single maturity model can fit all the various applications and contexts in which AI systems operate, we have found the TCMM to be a useful tool in such conversations.

Integration of the TCMM into AI system research and development efforts has already led to initial feedback on both the usefulness of the maturity model and the ways in which individual criteria were interpreted for specific applications. Future work could include user impact studies of the TCMM on calibrated user trust and the impact of that trust on user task performance. Such studies would ideally be conducted across various domains (e.g. Meier, 2025) and data modalities and include would experiments, surveys, or interviews with users, researchers and developers, and research directors or program managers. Future work could also expand on the domain-specific criteria details, providing a checklist-like scoring procedure for trust maturity for various subdomains.

AI systems are increasingly impactful across many applications of increasing consequence. In high consequence settings, dense technical information must be distilled to convey relevant implications to non-technical users. The TCMM is a tool for facilitating this process in a high-level package that contains the key information necessary for calibrating user trust.

## Declarations

Funding: The authors would like to acknowledge the US Department of Energy, National Nuclear Security Administration's Office of Defense Nuclear Nonproliferation Research and Development (NA-22) for supporting this work.

Conflict of interest: The authors declare that they have no known competing financial interests or personal relationships that could have appeared to influence the work reported in this paper.

Ethical approval: This research contains no elements requiring ethical approval.

Code and Data availability: We do not analyze or generate any code or datasets, because our work proceeds within a theoretical approach

## Bibliography

Acharya, A., Munikoti, S., Hellinger, A., Smith, S., Wagle, S., & Horawalavithana, S. (2023). *NuclearQA: A Human-Made Benchmark for Language Models for the Nuclear Domain* (No. arXiv:2310.10920). arXiv. https://doi.org/10.48550/arXiv.2310.10920

AI HLEG. (2019). High-level expert group on artificial intelligence. *Ethics Guidelines for Trustworthy AI*, *6*.

Amodei, D., Olah, C., Steinhardt, J., Christiano, P., Schulman, J., & Mané, D. (2016). *Concrete Problems in AI Safety* (No. arXiv:1606.06565). arXiv. https://doi.org/10.48550/arXiv.1606.06565

Bano, M., & Zowghi, D. (2015). A systematic review on the relationship between user involvement and system success. *Information and Software Technology*, *58*, 148–169. https://doi.org/10.1016/j.infsof.2014.06.011

Baroudi, J. J., Olson, M. H., & Ives, B. (1986). An empirical study of the impact of user involvement on system usage and information satisfaction. *Commun. ACM*, *29*(3), 232–238. https://doi.org/10.1145/5666.5669

Bommasani, R., Hudson, D. A., Adeli, E., Altman, R., Arora, S., von Arx, S., Bernstein, M. S., Bohg, J., Bosselut, A., Brunskill, E., Brynjolfsson, E., Buch, S., Card, D., Castellon, R., Chatterji, N., Chen, A., Creel, K., Davis, J. Q., Demszky, D., … Liang, P. (2022). *On the Opportunities and Risks of Foundation Models* (No. arXiv:2108.07258). arXiv. http://arxiv.org/abs/2108.07258

Bubeck, S., Chandrasekaran, V., Eldan, R., Gehrke, J., Horvitz, E., Kamar, E., Lee, P., Lee, Y. T., Li, Y., Lundberg, S., Nori, H., Palangi, H., Ribeiro, M. T., & Zhang, Y. (2023). *Sparks of Artificial General Intelligence: Early experiments with GPT-4* (No. arXiv:2303.12712). arXiv. http://arxiv.org/abs/2303.12712

Buchheit, M., Hirsch, F., & Martin, R. A. (2021). The Industrial Internet of Things trustworthiness framework foundations. *Industrial Internet Consortium. Available Online: Https://Www. Iiconsortium. Org/Pdf/Trustworthiness_Framework_Foundations. Pdf (Accessed on 30 December 2021)*.

Chai, C., Maceira, M., Santos-Villalobos, H. J., Venkatakrishnan, S. V., Schoenball, M., Zhu, W., Beroza, G. C., Thurber, C., & Team, E. C. (2020). Using a Deep Neural Network and Transfer Learning to Bridge Scales for Seismic Phase Picking. *Geophysical Research Letters*, *47*(16), e2020GL088651. https://doi.org/10.1029/2020GL088651

Chiang, W.-L., Zheng, L., Sheng, Y., Angelopoulos, A. N., Li, T., Li, D., Zhu, B., Zhang, H., Jordan, M., Gonzalez, J. E., & Stoica, I. (2024, June 6). *Chatbot Arena: An Open Platform for Evaluating LLMs by Human Preference*. Forty-first International Conference on Machine Learning. https://openreview.net/forum?id=3MW8GKNyzI

Chiou, E. K., & Lee, J. D. (2023). Trusting Automation: Designing for Responsivity and Resilience. *Human Factors*, *65*(1), 137–165. https://doi.org/10.1177/00187208211009995

Cho, S., Kim, I., Kim, J., Woo, H., & Shin, W. (2023). A Maturity Model for Trustworthy AI Software Development. *Applied Sciences*, *13*(8), 4771. https://doi.org/10.3390/app13084771

Corritore, C. L., Kracher, B., & Wiedenbeck, S. (2003). On-line trust: Concepts, evolving themes, a model. *International Journal of Human-Computer Studies*, *58*(6), 737–758. https://doi.org/10.1016/S1071-5819(03)00041-7

Das, S. D., Bala, P. K., & Mishra, A. N. (2024). Towards Defining a Trustworthy Artificial Intelligence System Development Maturity Model. *Journal of Computer Information Systems*, *64*(6), 775–796. https://doi.org/10.1080/08874417.2023.2251443

Dorton, S. L., & Harper, S. B. (2022). A Naturalistic Investigation of Trust, AI, and Intelligence Work. *Journal of Cognitive Engineering and Decision Making*, *16*(4), 222–236. https://doi.org/10.1177/15553434221103718

Goddard, K., Roudsari, A., & Wyatt, J. C. (2012). Automation bias: A systematic review of frequency, effect mediators, and mitigators. *Journal of the American Medical Informatics Association*, *19*(1), 121–127. https://doi.org/10.1136/amiajnl-2011-000089

Goodfellow, I. J., Shlens, J., & Szegedy, C. (2015). *Explaining and Harnessing Adversarial Examples* (No. arXiv:1412.6572). arXiv. https://doi.org/10.48550/arXiv.1412.6572

Goodwin, N. C. (1987). Functionality and usability. *Communications of the ACM*, *30*(3), 229–233. https://doi.org/10.1145/214748.214758

Gu, T., Dolan-Gavitt, B., & Garg, S. (2019). *BadNets: Identifying Vulnerabilities in the Machine Learning Model Supply Chain* (No. arXiv:1708.06733). arXiv. https://doi.org/10.48550/arXiv.1708.06733

Han, J., Kim, S., Sheen, D.-H., Lee, D., Lee, S.-J., Yoo, S.-H., & Park, D. (2023). Seismic event and phase detection using deep learning for the 2016 Gyeongju earthquake sequence. *Geosciences Journal*, *27*(3), 285–295. https://doi.org/10.1007/s12303-023-0004-y

Hancock, P. A., Billings, D. R., Schaefer, K. E., Chen, J. Y. C., de Visser, E. J., & Parasuraman, R. (2011). A Meta-Analysis of Factors Affecting Trust in Human-Robot Interaction. *Human Factors*, *53*(5), 517–527. https://doi.org/10.1177/0018720811417254

Hendrycks, D., Mazeika, M., & Woodside, T. (2023). *An Overview of Catastrophic AI Risks* (No. arXiv:2306.12001). arXiv. https://doi.org/10.48550/arXiv.2306.12001

Hoff, K. A., & Bashir, M. (2015). Trust in Automation: Integrating Empirical Evidence on Factors That Influence Trust. *Human Factors*, *57*(3), 407–434. https://doi.org/10.1177/0018720814547570

Hoffman, R. R., Mueller, S. T., Klein, G., & Litman, J. (2019). *Metrics for Explainable AI: Challenges and Prospects* (No. arXiv:1812.04608). arXiv. https://doi.org/10.48550/arXiv.1812.04608

*How to implement LLM guardrails | OpenAI Cookbook*. (n.d.). Retrieved April 7, 2025, from https://cookbook.openai.com/examples/how_to_use_guardrails

Hu, Y., Kuang, W., Qin, Z., Li, K., Zhang, J., Gao, Y., Li, W., & Li, K. (2021). Artificial Intelligence Security: Threats and Countermeasures. *ACM Comput. Surv.*, *55*(1), 20:1-20:36. https://doi.org/10.1145/3487890

*Human-AI Teaming: State-of-the-Art and Research Needs* (with National Academies of Sciences, Engineering, and Medicine). (2022). National Academies Press. https://doi.org/10.17226/26355

International Organization for Standardization (ISO). (2018). *Ergonomics of human-system interaction—Part 11: Usability: Definitions and concepts (ISO 9241-11)*. https://www.iso.org/standard/63500.html

Jacovi, A., Marasović, A., Miller, T., & Goldberg, Y. (2021). Formalizing Trust in Artificial Intelligence: Prerequisites, Causes and Goals of Human Trust in AI. *Proceedings of the 2021 ACM Conference on Fairness, Accountability, and Transparency*, 624–635. https://doi.org/10.1145/3442188.3445923

Kujala, S. (2003). User involvement: A review of the benefits and challenges. *Behaviour & Information Technology*. https://doi.org/10.1080/01449290301782

*Lessons learned on language model safety and misuse*. (2024, February 14). https://openai.com/index/language-model-safety-and-misuse/

Liang, P., Bommasani, R., Lee, T., Tsipras, D., Soylu, D., Yasunaga, M., Zhang, Y., Narayanan, D., Wu, Y., Kumar, A., Newman, B., Yuan, B., Yan, B., Zhang, C., Cosgrove, C., Manning, C. D., Ré, C., Acosta-Navas, D., Hudson, D. A., … Koreeda, Y. (2023). *Holistic Evaluation of Language Models* (No. arXiv:2211.09110). arXiv. https://doi.org/10.48550/arXiv.2211.09110

Liao, Q. V., & Vaughan, J. W. (2023). AI transparency in the age of LLMs: A human-centered research roadmap. *arXiv Preprint arXiv:2306.01941*, *10*.

Lu, P., Mishra, S., Xia, T., Qiu, L., Chang, K.-W., Zhu, S.-C., Tafjord, O., Clark, P., & Kalyan, A. (2022). Learn to Explain: Multimodal Reasoning via Thought Chains for Science Question Answering. *Advances in Neural Information Processing Systems*, *35*, 2507–2521.

Mankins, J. C. (1995). Technology readiness levels. *White Paper*.

Mayer, R. C., Davis, J. H., & Schoorman, F. D. (1995). An Integrative Model of Organizational Trust. *Academy of Management Review*, *20*(3), 709–734. https://doi.org/10.5465/amr.1995.9508080335

Meier, R. (2025). Balancing Minds and Data: The Privacy Dilemma of LLMs and Anthropomorphism in LLMs. *Journal of Social Computing*, *6*(3), 173–183. https://doi.org/10.23919/JSC.2025.0014

Michelini, A., Cianetti, S., Gaviano, S., Giunchi, C., Jozinović, D., & Lauciani, V. (2021). INSTANCE – the Italian seismic dataset for machine learning. *Earth System Science Data*, *13*(12), 5509–5544. https://doi.org/10.5194/essd-13-5509-2021

Miller, C. A. (2021). Chapter 11 - Trust, transparency, explanation, and planning: Why we need a lifecycle perspective on human-automation interaction. In C. S. Nam & J. B. Lyons (Eds.), *Trust in Human-Robot Interaction* (pp. 233–257). Academic Press. https://doi.org/10.1016/B978-0-12-819472-0.00011-3

Mishra, A. (2018). Metrics to evaluate your machine learning algorithm. *Towards Data Science*, *24*. https://medium.com/data-science/metrics-to-evaluate-your-machine-learning-algorithm-f10ba6e38234

Münchmeyer, J., Woollam, J., Rietbrock, A., Tilmann, F., Lange, D., Bornstein, T., Diehl, T., Giunchi, C., Haslinger, F., Jozinović, D., Michelini, A., Saul, J., & Soto, H. (2022). Which Picker Fits My Data? A Quantitative Evaluation of Deep Learning Based Seismic Pickers. *Journal of Geophysical Research: Solid Earth*, *127*(1), e2021JB023499. https://doi.org/10.1029/2021JB023499

Myren, S., Parikh, N., Rael, R., Flynn, G., Higdon, D., & Casleton, E. (2025). Evaluation of Seismic Artificial Intelligence with Uncertainty. *Seismological Research Letters*. https://doi.org/10.1785/0220240444

Nielsen, J. (1994). *Usability Engineering*. Morgan Kaufmann.

Nielsen, J. (2012). How many test users in a usability study. *Nielsen Norman Group*, *4*(06).

Oberkampf, W. L., Trucano, T. G., & Pilch, M. M. (2007). *Predictive Capability Maturity Model for computational modeling and simulation.* (Nos. SAND2007-5948). Sandia National Laboratories (SNL), Albuquerque, NM, and Livermore, CA (United States). https://doi.org/10.2172/976951

*OpenAI safety practices*. (2024). https://openai.com/index/openai-safety-update/

Papenmeier, A., Englebienne, G., & Seifert, C. (2019). *How model accuracy and explanation fidelity influence user trust* (No. arXiv:1907.12652). arXiv. https://doi.org/10.48550/arXiv.1907.12652

Parasuraman, R., & Manzey, D. H. (2010). Complacency and Bias in Human Use of Automation: An Attentional Integration. *Human Factors*, *52*(3), 381–410. https://doi.org/10.1177/0018720810376055

Parasuraman, R., & Riley, V. (1997). Humans and Automation: Use, Misuse, Disuse, Abuse. *Human Factors*, *39*(2), 230–253. https://doi.org/10.1518/001872097778543886

Park, Y., Delbridge, B. G., & Shelly, D. R. (2024). Making Phase-Picking Neural Networks More Consistent and Interpretable. *The Seismic Record*, *4*(1), 72–80. https://doi.org/10.1785/0320230054

Paszke, A., Gross, S., Massa, F., Lerer, A., Bradbury, J., Chanan, G., Killeen, T., Lin, Z., Gimelshein, N., Antiga, L., Desmaison, A., Köpf, A., Yang, E., DeVito, Z., Raison, M., Tejani, A., Chilamkurthy, S., Steiner, B., Fang, L., … Chintala, S. (2019). *PyTorch: An Imperative Style, High-Performance Deep Learning Library* (No. arXiv:1912.01703). arXiv. http://arxiv.org/abs/1912.01703

Paulk, M. C., Curtis, B., Chrissis, M. B., & Weber, C. V. (1993). Capability maturity model, version 1.1. *IEEE Software*, *10*(4), 18–27. https://doi.org/10.1109/52.219617

Qi, X., Huang, Y., Zeng, Y., Debenedetti, E., Geiping, J., He, L., Huang, K., Madhushani, U., Sehwag, V., Shi, W., Wei, B., Xie, T., Chen, D., Chen, P.-Y., Ding, J., Jia, R., Ma, J., Narayanan, A., Su, W. J., … Mittal, P. (2024). *AI Risk Management Should Incorporate Both Safety and Security* (No. arXiv:2405.19524). arXiv. http://arxiv.org/abs/2405.19524

Quiring, E., Klein, D., Arp, D., Johns, M., & Rieck, K. (2020). Adversarial Preprocessing: Understanding and Preventing Image-Scaling Attacks in Machine Learning. *Proceedings of the 29th USENIX Security Symposium*.

Radosevich, B., & Halloran, J. (2025). *MCP Safety Audit: LLMs with the Model Context Protocol Allow Major Security Exploits* (No. arXiv:2504.03767). arXiv. https://doi.org/10.48550/arXiv.2504.03767

Ray, P. P. (2023). ChatGPT: A comprehensive review on background, applications, key challenges, bias, ethics, limitations and future scope. *Internet of Things and Cyber-Physical Systems*, *3*, 121–154. https://doi.org/10.1016/j.iotcps.2023.04.003

Ronneberger, O., Fischer, P., & Brox, T. (2015). U-Net: Convolutional Networks for Biomedical Image Segmentation. In N. Navab, J. Hornegger, W. M. Wells, & A. F. Frangi (Eds.), *Medical Image Computing and Computer-Assisted Intervention – MICCAI 2015* (pp. 234–241). Springer International Publishing. https://doi.org/10.1007/978-3-319-24574-4_28

Schaschek, M., & Engel, S. (2023). *Measuring Trustworthiness of AI Systems: A Holistic Maturity Model*.

Sheridan, T. B. (2019). Extending Three Existing Models to Analysis of Trust in Automation: Signal Detection, Statistical Parameter Estimation, and Model-Based Control. *Human Factors*, *61*(7), 1162–1170. https://doi.org/10.1177/0018720819829951

Smith, M., Martinez, C., Ingram, J., DeBonis, M., Cuellar, C., & Jose, D. (2023). Test and Evaluation of Systems with Embedded Machine Learning Components. *The ITEA Journal of Test and Evaluation*, *44*(3). https://itea.org/journals/volume-44-3/test-and-evaluation-of-systems-with-embedded-ml-components/

Smith, M. R., Acquesta, E., Smutz, C., Rushdi, A., & Moss, B. (2022). *Assessing the Fidelity of Explanations with Global Sensitivity Analysis* (Nos. SAND2022-12331C). Sandia National Lab. (SNL-NM), Albuquerque, NM (United States). https://www.osti.gov/servlets/purl/2004705

Speed, A., & Stracuzzi, D. J. (2020). *Research Needs for Trusted Analytics in National Security Settings* (No. SAND-2020-9709R). Sandia National Lab. (SNL-NM), Albuquerque, NM (United States). https://doi.org/10.2172/1662023

Stanton, B., & Jensen, T. (2021). Trust and Artificial Intelligence (Draft). *NIST*. https://www.nist.gov/publications/trust-and-artificial-intelligence-draft

Stites, M. C., Nyre-Yu, M., Moss, B., Smutz, C., & Smith, M. R. (2021). Sage Advice? The Impacts of Explanations for Machine Learning Models on Human Decision-Making in Spam Detection. In H. Degen & S. Ntoa (Eds.), *Artificial Intelligence in HCI* (pp. 269–284). Springer International Publishing. https://doi.org/10.1007/978-3-030-77772-2_18

Tabassi, E. (2023). Artificial Intelligence Risk Management Framework (AI RMF 1.0). *NIST*. https://www.nist.gov/publications/artificial-intelligence-risk-management-framework-ai-rmf-10

Templeton, A. (2024). *Scaling monosemanticity: Extracting interpretable features from claude 3 sonnet*. Anthropic.

*Trustworthy Artificial Intelligence (AI)TM*. (2020). https://www2.deloitte.com/us/en/pages/deloitte-analytics/solutions/ethics-of-ai-framework.html

Tustison, N. J., Johnson, H. J., Rohlfing, T., Klein, A., Ghosh, S. S., Ibanez, L., & Avants, B. (2013). Instrumentation bias in the use and evaluation of scientific software: Recommendations for reproducible practices in the computational sciences. *Frontiers in Neuroscience*, *7*. https://doi.org/10.3389/fnins.2013.00162

Wang, J., Hu, X., Hou, W., Chen, H., Zheng, R., Wang, Y., Yang, L., Huang, H., Ye, W., Geng, X., Jiao, B., Zhang, Y., & Xie, X. (2023). *On the Robustness of ChatGPT: An Adversarial and Out-of-distribution Perspective* (No. arXiv:2302.12095). arXiv. https://doi.org/10.48550/arXiv.2302.12095

Wei, J., Wang, X., Schuurmans, D., Bosma, M., Ichter, B., Xia, F., Chi, E., Le, Q. V., & Zhou, D. (2022). Chain-of-Thought Prompting Elicits Reasoning in Large Language Models. *Advances in Neural Information Processing Systems*, *35*, 24824–24837.

Zhu, W., & Beroza, G. C. (2018). PhaseNet: A Deep-Neural-Network-Based Seismic Arrival Time Picking Method. *Geophysical Journal International*. https://doi.org/10.1093/gji/ggy423